\documentclass[10pt,twocolumn,letterpaper]{article}

\usepackage{cvpr}              

\usepackage{multirow}
\usepackage{multicol}
\usepackage{amsmath}
\usepackage{makecell}
\usepackage{float}
\usepackage{amssymb}
\usepackage{booktabs}
\usepackage{subcaption}
\usepackage{xcolor,colortbl}
\DeclareMathOperator*{\argmax}{arg\,max}
 
\usepackage{bm} 
\usepackage{graphicx}
\usepackage{amstext}

\definecolor{cvprblue}{rgb}{0.21,0.49,0.74}
\usepackage[pagebackref,breaklinks,colorlinks,allcolors=cvprblue]{hyperref}

\def\paperID{} 
\def\confName{CVPR}
\def\confYear{2025}

\newcommand{\fc}[1]{
\ifthenelse{\equal{\commentisunseen}{0}}{
{\color{blue}#1}}
{#1}\xspace
}
\definecolor{lblue}{RGB}{70,130,180} 
\newcommand{\blue}[1]{\textcolor{blue}{\textbf{#1}}}
\newcommand{\lblue}[1]{\textcolor{lblue}{\textbf{#1}}}
\title{CodeMMR: Bridging Natural Language, Code, and Image for Unified Retrieval}

\author{
Jiahui Geng$^{1,2}$ \quad
Qing Li$^{1,3}$\thanks{Corresponding author: qing.li@rug.nl} \quad
Fengyu Cai$^{4}$ \quad
Fakhri Karray$^{1}$
\\
$^{1}$MBZUAI, UAE \quad
$^{2}$Linköping University, Sweden \\
$^{3}$University of Groningen, Netherlands \quad
$^{4}$TU Darmstadt, Germany
}

\begin{document}
\maketitle

\begin{abstract}
Code search, framed as information retrieval (IR), underpins modern software engineering and increasingly powers retrieval-augmented generation (RAG), improving code discovery, reuse, and the reliability of LLM-based coding.
Yet existing code IR models remain largely text-centric and often overlook the visual and structural aspects inherent in programming artifacts such as web interfaces, data visualizations, SVGs, schematic diagrams, and UML.
To bridge this gap, we introduce \textbf{MMCoIR}, the first comprehensive benchmark for evaluating multimodal code IR across \textbf{five} visual domains, \textbf{eight} programing languages, \textbf{eleven} libraries,  and show the challenge of the task through extensive evaluation.
Therefore, we then propose \textbf{CodeMMR}, a unified retrieval model that jointly embeds natural language, code, and images into a shared semantic space through instruction-based multimodal alignment.
CodeMMR achieves strong generalization across modalities and languages, outperforming competitive baselines (e.g., UniIR, GME, VLM2Vec) by an average of 10 points on nDCG@10.
Moreover, integrating CodeMMR into RAG enhances code generation fidelity and visual grounding on unseen code generation tasks, underscoring the potential of multimodal retrieval as a core enabler for next-generation intelligent programming systems. Datasets are available at HuggingFace\footnote{\url{https://huggingface.co/datasets/JiahuiGengNLP/MMCoIR-train}}$^{,}$\footnote{\url{https://huggingface.co/datasets/JiahuiGengNLP/MMCoIR-test}}.
\end{abstract}

\section{Introduction}

Code search, typically implemented via information retrieval (IR), is a cornerstone of modern software engineering, enabling developers to effectively locate, understand, and reuse relevant code snippets~\cite{sachdev2018retrieval,husain2019codesearchnet,li-etal-2025-coir,geng2025coquir}.
Code IR can further improves software quality by surfacing related explanations, bug analyses, and design patterns,as evidenced by its integration into commercial products such as Commercial products such as GitHub Code Search and VS Code’s intelligent lookup tools~\cite{del2023introducing}.
Recently, code retrieval-augmented generation (RAG) systems~\cite{su-etal-2024-evor,zhang-etal-2023-repocoder,zhang-etal-2023-syntax} have leveraged retrieval to mitigate hallucinations in large language models (LLMs), improving reliability and grounding in programming-related tasks, such as code generation.

Despite advances in code IR, existing systems remain predominantly text-based, focusing on semantic similarity between code and natural language~\cite{husain2019codesearchnet,li-etal-2025-coir,geng2025coquir}. In practice, however, modern software artifacts are deeply multimodal: code often embodies visual functionality, such as defining a web layout~\cite{yun2024web2code,gui2025webcode2m}, rendering a chart~\cite{kondic2025chartgen,yang2025chartmimic,zhao-etal-2025-chartcoder}, or generating a UML diagram~\cite{bates2025unified}, etc. 
Developers need to understand what a piece of code looks like when executed, or, conversely, refer to the code that can generate a given visual result from an image or design.
This mismatch highlights an urgent need for multimodality in code IR.

\begin{figure*}
    \centering
    \includegraphics[width=0.95\linewidth]{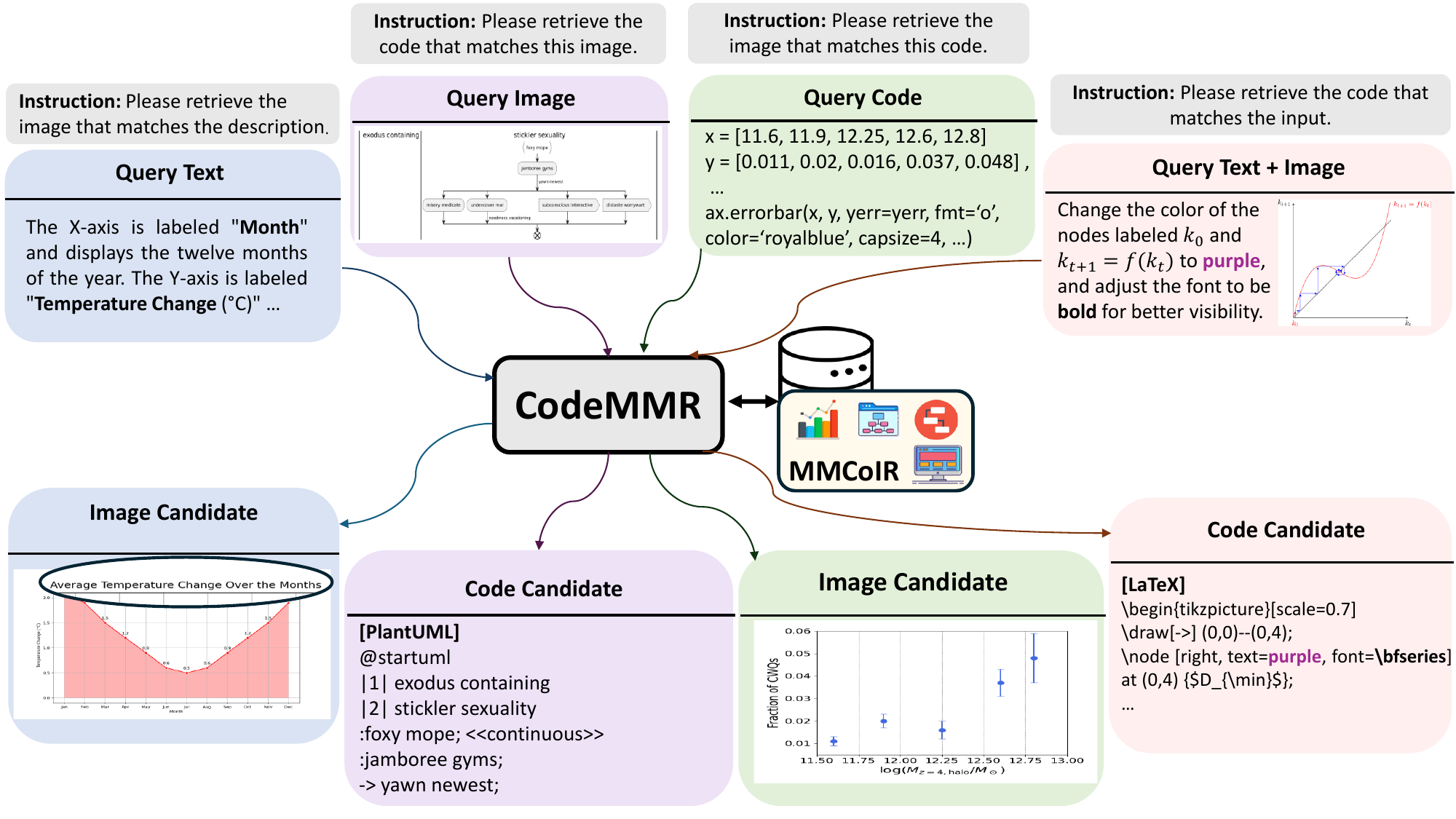}
    \caption{\textbf{Illustration of CodeMMR and MMCoIR}: Unified multimodal CodeMMR for text, image, and code, evaluated on MMCoIR.}
    \label{fig:placeholder}
\end{figure*}

To bridge this gap, we introduce \textbf{MMCoIR}, a comprehensive benchmark for systematically evaluating \emph{multimodal multilingual code retrieval} models across diverse domains. As shown in Table~\ref{tab:datasets}, MMCoIR unifies datasets from \emph{WebUI}, \emph{data charts}, \emph{SVGs}, \emph{schematic diagrams}, and \emph{UML diagrams}, each containing paired samples of images and codes in \emph{eight} programming languages (e.g., \emph{HTML, CSS, JS, Python, XML, LaTeX, PlantUML, etc}) and \emph{eleven} libraries. This design supports a wide range of retrieval tasks, including \emph{text-to-code}, \emph{image-to-code}, and \emph{text+image-to-code}, as well as their reverse directions for visual retrieval. The textual queries include both natural image descriptions and instruction-style prompts derived from code editing and repair datasets. As the first \textbf{multimodal} and \textbf{multilingual} code retrieval testbed, MMCoIR enables research for more realistic and diverse retrieval scenarios.

Multimodal code retrieval is challenging;
we comprehensively evaluating the existing multimodal embedding models~\cite{zhang2024gme,jiang2025vlmvec,liu2025lamra} spanning different architectures and scales on MMCoIR and finding their performance lacking despite strong model initialization and large-scale training.
Therefore, we introduce \textbf{CodeMMR}, a unified multimodal, multilingual code retrieval model that jointly encodes \textbf{natural language}, \textbf{code}, and \textbf{visual} modalities into a shared embedding space.
Trained through instruction-based multimodal alignment, CodeMMR delivers strong \emph{in-} and \emph{out-of-}distribution perforamnce across all MMCoIR subtasks, surpassing VLM2Vec-v2~\cite{meng2025vlm2vec} by an average of \textbf{10} absolute points on nDCG@10, which demonstrates its accurate and robust retrieval across code, image, and text.
On two unseen image-to-code generation tasks, ChartMimic Direct and WebCode2M-Mid, RAG with CodeMMR outperforms the non-retrieval baseline, by 10.0 points in Execution Rate and 9.4 points in Visual Accuracy, respectively, and also beats RAG with baseline retrievers, demonstrating its substantial practical values.
Moreover, we outline promising directions for multimodal code retrieval, including long-context retrieval (e.g., SVG) and fine-grained retrieval along specific dimensions, e.g., text, layout, and color.
Overall, our contributions are threefold:
\begin{itemize}
    \item We introduce \textbf{MMCoIR}, the first large-scale benchmark for embedding model evaluation on \emph{multimodal multilingual code retrieval} across domains and modalities.
    \item We further train \textbf{CodeMMR}, a unified embedding model that aligns text, code, and visual representations through instruction-based multimodal training.
    \item Extensive experiments show that CodeMMR outperforms existing baselines and improves RAG-based code generation, showcasing significant practical value.
\end{itemize}

\section{Related Work}

\paragraph{Multimodal Embedding and Universal Representation Learning} Early vision–language models (VLMs) such as CLIP~\cite{pmlr-v139-radford21a}, ALIGN~\cite{jia2021scaling}, and BLIP~\cite{li2022blip} demonstrated that large-scale contrastive learning between image–text pairs can produce general-purpose multimodal representations.
Subsequent research extended this idea toward universal multimodal embeddings capable of handling diverse downstream tasks within a shared semantic space~\cite{wei2024uniir,10.5555/3692070.3694524,jiang2025vlmvec,meng2025vlm2vec}.
Frameworks like UniIR~\cite{wei2024uniir} and MagicLens~\cite{10.5555/3692070.3694524} unified multiple retrieval tasks through shared encoders and contrastive objectives, showing improved task generalization.
MM-EMBED~\cite{lin2025mmembed} first unified text, image, and video representations via multimodal large language models (MLLMs) under an instruction-driven retrieval objective.
Building on this, VLM2Vec~\cite{jiang2025vlmvec} advanced vision–language models through instruction-conditioned contrastive learning, enabling scalable multitask multimodal embedding within a shared representation space. Further advances such as Think-then-Embed~\cite{cui2025think} incorporate generative reasoning contexts prior to embedding, enhancing compositional understanding and semantic coherence. LamRA~\cite{liu2025lamra} leveraged large multimodal models as retrieval assistants, bridging encoder-based embedding and cross-modal reasoning via adaptive attention integration. Overall, recent advances point toward unifying multimodal retrieval with MLLM-based embeddings; here we take the next step by incorporating code, a critical yet long-overlooked modality.

\begin{table*}[htbp]
\centering
\small
\setlength{\tabcolsep}{3pt}
\resizebox{0.85\textwidth}{!}{%
\begin{tabular}{l|>{\raggedright\arraybackslash}m{2.2cm}|>{\raggedright\arraybackslash}m{2.8cm}|>{\raggedright\arraybackslash}m{2.8cm}|p{2.5cm}|p{2.5cm}}
\hline
\textbf{Domains} & \textbf{Datasets} & \textbf{Images} & \textbf{Languages} & \textbf{\#Training} & \textbf{\#Test} \\ \hline
\multirow{4}{*}{WebUI} 
& \multirow{2}{=}{WebSight~\cite{laurenccon2024unlocking}} & \multirow{2}{=}{Screenshots} & \multirow{2}{=}{HTML+CSS+JS} & $q_i \rightarrow r_c$ (100k) & $q_i \rightarrow r_c$ (100k) \\
& & & & $q_c \rightarrow r_i$ (100k) & $q_c \rightarrow r_i$ (100k) \\ \cline{2-6}
& Web2Code~\cite{yun2024webcode} & Screenshots & HTML+CSS+JS & $q_i \rightarrow r_c$ (100k) & $q_i \rightarrow r_c$ (2k) \\ \cline{2-6}
& Sketch2Code~\cite{li-etal-2025-sketch2code} & Hand-drawn sketches & & & $q_i \rightarrow r_c$ (2k) \\ \hline
\multirow{9}{*}{\shortstack[l]{Data\\Charts}}  
& \multirow{2}{=}{Chart2Code~\cite{zhao-etal-2025-chartcoder}} & \multirow{2}{=}{Chart images} & \multirow{2}{=}{Python (2 libraries)} & $q_i \rightarrow r_c$ (100k) & $q_i \rightarrow r_c$ (2k) \\ 
& & & & & \\ \cline{2-6}
& \multirow{4}{=}{ChartGen~\cite{kondic2025chartgen}} & \multirow{4}{=}{Chart images} & \multirow{4}{=}{Python (11~libraries)} & $q_i \rightarrow r_c$ (100k) & $q_i \rightarrow r_c$ (2k) \\ 
& & & & $q_c \rightarrow r_i$ (100k) & $q_c \rightarrow r_i$ (2k) \\ 
& & & & $q_t \rightarrow r_c$ (100k) & $q_t \rightarrow r_c$ (2k) \\ 
& & & & $q_t \rightarrow r_i$ (100k) & $q_t \rightarrow r_i$ (2k) \\ \cline{2-6}
& \multirow{4}{=}{ChartEdit~\cite{zhao-etal-2025-chartedit}} & \multirow{4}{=}{Chart images} & \multirow{4}{=}{Python (2~libraries)} & & $q_i \rightarrow r_c$ (2k) \\ 
& & & & & $q_c \rightarrow r_i$ (2k) \\ 
& & & & & $q_{t,c} \rightarrow r_i$ (2k) \\ 
& & & & & $q_{t,c} \rightarrow r_{i,c}$ (2k) \\ \hline
\multirow{6}{*}{SVG} 
& \multirow{2}{=}{SVGStack~\cite{Rodriguez_2025_CVPR}} & \multirow{2}{=}{SVG images} & \multirow{2}{=}{XML for SVG} & $q_i \rightarrow r_c$ (100k) & $q_i \rightarrow r_c$ (2k) \\
& & & & $q_c \rightarrow r_i$ (100k) & $q_c \rightarrow r_i$ (12k) \\ \cline{2-6}
& \multirow{4}{=}{MMSVG~\cite{yang2025omnisvg}} & \multirow{4}{=}{SVG images} & \multirow{4}{=}{XML for SVG} & $q_i \rightarrow r_c$ (200k) & $q_i \rightarrow r_c$ (4k) \\
& & & & $q_c \rightarrow r_i$ (200k) & $q_c \rightarrow r_i$ (4k) \\
& & & & $q_t \rightarrow r_c$ (200k) & $q_t \rightarrow r_c$ (4k) \\
& & & & $q_t \rightarrow r_i$ (200k) & $q_t \rightarrow r_i$ (4k) \\ \hline
\multirow{6}{*}{\shortstack[l]{Schematic\\Diagrams}} 
& \multirow{4}{=}{\shortstack[l]{DiagramGen\\Benchmark~\cite{wei2025words}}} & \multirow{4}{=}{Diagrams} & \multirow{4}{=}{\shortstack[l]{LaTeX + TikZ,\\Graphviz DOT}} & & $q_i \rightarrow r_c$ (270) \\
& & & & & $q_c \rightarrow r_i$ (270) \\
& & & & & $q_t \rightarrow r_c$ (267) \\
& & & & & $q_{t,i} \rightarrow r_c$ (200) \\ \cline{2-6}
& \multirow{2}{=}{DATIKZ$_{v3}$~\cite{belouadi2025tikzero}} & \multirow{2}{=}{Diagrams} & \multirow{2}{=}{LaTeX + TikZ} & $q_i \rightarrow r_c$ (100k) & $q_i \rightarrow r_c$ (2k) \\
& & & & $q_c \rightarrow r_i$ (100k) & $q_c \rightarrow r_i$ (2k) \\ \hline
\multirow{2}{*}{SE UML} & \multirow{2}{=}{PlantUML~\cite{bates2025unified}} & \multirow{2}{=}{UML diagrams} & \multirow{2}{=}{PlantUML scripts} & $q_i \rightarrow r_c$ (200k) & $q_i \rightarrow r_c$ (4k) \\
& & & & $q_c \rightarrow r_i$ (200k) & $q_c \rightarrow r_i$ (4k) \\ \hline
\end{tabular}%
}
\caption{\textbf{Overview of MMCoIR}, which covers various \emph{domains}, \emph{languages}, \emph{libraries}, and \emph{format combinations} of queries and targets.}
\label{tab:datasets}
\end{table*}

\paragraph{Code IR and RAG}
Early datasets such as CodeSearchNet~\cite{husain2019codesearchnet} and CoSQA~\cite{huang-etal-2021-cosqa} enabled text–code semantic alignment evaluation but lacked domain and task diversity.
CoIR~\cite{li-etal-2025-coir} addressed this by unifying ten datasets and eight retrieval types across fourteen programming languages; CoQuIR~\cite{geng2025coquir} further enriched it with quality-aware annotations, and XCodeEval~\cite{khan-etal-2024-xcodeeval} extended evaluation to multilingual, multi-task settings.
On the methodological side, CodeXEmbed~\cite{liu2025codexembed} applies a two-stage training paradigm to adapt general-purpose retrievers to code domains, while Revela~\cite{cai2025revela} achieves competitive CoIR performance through unsupervised retriever learning framed as language modeling.
In generation, code RAG~\cite{wang-etal-2025-coderag} reduces hallucinations by grounding code generation with retrieved reference code, documentation, and tutorials.
Yet existing work treats code purely as text, overlooking visual modalities—such as web interfaces and visualizations—that are integral to real-world human–machine interaction.
We address this gap by extending code retrieval and RAG to incorporate these code-related visual modalities.

\section{Task Overview}

\subsection{Problem: Code Multimodal Retrieval}
\label{sec:problem_definition}


In a unified multimodal code retrieval framework, users issue queries originating from diverse modalities, including natural language, source code, and visual content.
Formally, a query is denoted as $\mathbf{q}$ and may take the form of text $q_{\text{t}}$, image $q_{\text{i}}$, or code $q_{\text{c}}$. 
Beyond unimodal cases, certain retrieval scenarios involve \textit{composed queries} that combine multiple modalities (e.g., $q_{\text{t,i}}$ or $q_{\text{t,c}}$), which are prevalent in tasks such as code editing, code repair, and visualization refinement~\cite{zhao-etal-2025-chartedit,wei2025words}. 
Likewise, a retrieval target $\mathbf{r}$ can correspond to any single or composite modality, such as $r_{\text{i}}$, $r_{\text{c}}$, or $r_{\text{i,c}}$. 
This formulation allows a single retriever to accommodate heterogeneous search paradigms, such as text-to-code, image-to-code, and text+image-to-code, under a unified representation framework, as summarized in Table~\ref{tab:datasets}.

To ensure flexibility of retrieval intent across heterogeneous tasks, each query is accompanied by a natural-language instruction $q_{\text{inst}}$, which explicitly specifies the retrieval goal and domain context (e.g., ``please retrieve the code that matches this image.'').  
This instruction-based design clarifies task semantics and improves compatibility across datasets with varying modality support.  
For example, datasets containing only image--code pairs ($q_{\text{i}} \rightarrow r_{\text{c}}$ or $q_{\text{c}} \rightarrow r_{\text{i}}$) are augmented with standardized prompts to indicate the intended retrieval direction.  
In more complex composed scenarios, such as $q_{\text{t,i}} \rightarrow r_{\text{c}}$, the textual component $q_{\text{t}}$ is typically drawn from dataset-provided prompts (e.g., ``change the color of the nodes to purple.'').

The retrieval process is then defined as follows: given a query $\mathbf{q}$ and its corresponding instruction $q_{\text{inst}}$, the goal is to identify the most relevant candidate $r^{\ast}$ from a heterogeneous candidate pool $\mathcal{R}$ by maximizing the similarity in a shared multimodal embedding space:
\begin{align*}
    r^{\ast} = \argmax_{\mathbf{r} \in \mathcal{R}} 
    \left[ f_{\theta}(\mathbf{q}, q_{\text{inst}})^{\top} f_{\theta}(\mathbf{r}) \right],
\end{align*}
where $f_{\theta}(\cdot)$ denotes a multimodal encoder parameterized by $\theta$.  
This unified formulation enables instruction-conditioned retrieval across natural language, code, and visual domains.  
A detailed description of the datasets and retrieval configurations is provided in Section~\ref{sec:mmcoir}, where we introduce the proposed benchmark, \textbf{MMCoIR}.

\subsection{Solution: Code Multimodal Retriever Training}
\label{sec:training_for_retrieval}
To enable effective cross-modal retrieval between natural language, code, and images, we train \textbf{CodeMMR} to project heterogeneous modalities into a shared embedding space.  
Given a pretrained vision–language model (VLM) as the initialization, CodeMMR extends its capability to understand structured code representations and to align them with both textual and visual semantics.

\vspace{2pt}\noindent
\textbf{Training Objective.}  
We employ a contrastive learning objective based on the InfoNCE loss~\cite{oord2018representation}.  
For a batch of $B$ query–target pairs $\{(\mathbf{q}_i, \mathbf{r}_i^+)\}_{i=1}^{B}$, CodeMMR encodes each instruction-conditioned query and target as:
\[
\mathbf{h}_{q_i} = f_{\theta}(\mathbf{q}_i, q_{\text{inst}}), \quad 
\mathbf{h}_{r_i^+} = f_{\theta}(\mathbf{r}_i^+),
\]
where $f_{\theta}$ is the multimodal encoder parameterized by $\theta$.  
The model maximizes the similarity between each query and its positive target while contrasting against all negatives $\mathbf{r}_i^- \in \mathcal{N}$.  
The loss is defined as:
\begin{equation}
\label{equ:infoNCE_codemmr}
\mathcal{L}_{\text{ret}} = -\frac{1}{B} \sum_{i=1}^{B} 
\log \frac{\phi(\mathbf{h}_{q_i}, \mathbf{h}_{r_i^+})}
{\phi(\mathbf{h}_{q_i}, \mathbf{h}_{r_i^+}) + \sum_{\mathbf{r}_i^- \in \mathcal{N}} \phi(\mathbf{h}_{q_i}, \mathbf{h}_{r_i^-})},
\end{equation}
where $\phi(\mathbf{h}_q, \mathbf{h}_r) = \exp\!\left(\frac{1}{\tau}\mathbf{h}_q^{\top}\mathbf{h}_r\right)$  
is a temperature-scaled similarity and $\tau$ is a hyper-parameter.  
We incorporate both in-batch negatives and hard negatives drawn from semantically close examples to improve discriminative alignment.

\section{MMCoIR Benchmark}
\label{sec:mmcoir}

\subsection{Overview}
\label{sec:dataset_collection}

The MMCoIR benchmark establishes a unified multimodal code retrieval framework for comprehensive evaluation across five representative visual domains: (i) web development and UI design, (ii) data charts, (iii) scalable vector graphics, (iv) schematic diagrams, and (v) software engineering (UML).
It unifies heterogeneous datasets across multiple programming languages and visual modalities, enabling systematic evaluation of cross-domain and cross-lingual understanding. Specifically, MMCoIR incorporates datasets from several key areas:

\begin{itemize}
\item \textbf{Web Development}: WebSight~\cite{laurenccon2024unlocking}, Web2Code~\cite{yun2024webcode}, and Sketch2Code~\cite{li-etal-2025-sketch2code} translate realistic webpage screenshots into HTML/CSS code, emphasizing layout understanding and visual fidelity.
\item \textbf{Data Visualization}: Chart2Code~\cite{zhao-etal-2025-chartcoder}, ChartGen~\cite{kondic2025chartgen}, and ChartEdit~\cite{zhao-etal-2025-chartedit} capture the correspondence between data and visual encodings.
\item \textbf{Vector Graphics}: SVGStack~\cite{Rodriguez_2025_CVPR} and MMSVG~\cite{yang2025omnisvg} involve structured SVG representations for scalable vector graphic synthesis and manipulation.
\item \textbf{Schematic Diagrams}: DiagramGenBenchmark~\cite{wei2025words} and DATIKZ$_{v3}$~\cite{belouadi2025tikzero} focus on schematic or symbolic diagrams with LaTeX~TikZ and Graphviz-style languages.
\item \textbf{Software Engineering}: PlantUML~\cite{bates2025unified} targets software diagrams such as class and sequence diagrams, highlighting semantic precision and logical relationships.
\end{itemize}

Together, they cover a linguistically diverse spectrum—from markup and scripting languages (HTML, CSS, JavaScript, XML, Python) to domain-specific formats such as LaTeX TikZ and PlantUML.
Overall, MMCoIR provides the first comprehensive foundation for evaluating multimodal and multilingual code retrieval systems, fostering research toward more generalizable and semantically grounded cross-modal embeddings. Please refer to supplmentary documents for more details.

\begin{table*}[t]
\centering
\setlength{\tabcolsep}{2pt}
\resizebox{\textwidth}{!}{
\begin{tabular}{lcc|c|c|cc|cc|cccc|cc|cc|c}
\toprule
\multirow{3}{*}{\textbf{\large Methods}} & 
\multicolumn{3}{c|}{\textbf{\large WebUI}} & 
\multicolumn{3}{c|}{\textbf{\large Data Charts}} & 
\multicolumn{6}{c|}{\textbf{\large SVG}} & 
\multicolumn{2}{c|}{\textbf{\large Schematic}} & 
\multicolumn{2}{c|}{\textbf{\large UML }} &
\multirow{3}{*}{\textbf{\large Avg}} \\
\cmidrule(lr){2-17}
& \multicolumn{2}{c|}{\textbf{WebSight}} & 
\textbf{Web2Code} & 
\textbf{Chart2Code} & 
\multicolumn{2}{c|}{\textbf{ChartGen}} & 
\multicolumn{2}{c|}{\textbf{SVGStack}} & 
\multicolumn{4}{c|}{\textbf{MMSVG}} & 
\multicolumn{2}{c|}{\textbf{DATIKZv3}} & 
\multicolumn{2}{c|}{\textbf{PlantUML}} & \\
\cmidrule(lr){2-17}
& $q_c \to r_i$ & $q_i \to r_c$ & 
$q_i \to r_c$ & 
$q_i \to r_c$ & 
$q_c \to r_i$ & $q_i \to r_c$ & 
$q_c \to r_i$ & $q_i \to r_c$ & 
$q_c \to r_i$ & $q_i \to r_c$ & $q_t \to r_c$ & $q_t \to r_i$ & 
$q_c \to r_i$ & $q_i \to r_c$ & 
$q_c \to r_i$ & $q_i \to r_c$ & \\
\midrule
\multicolumn{18}{c}{\textit{Metric: Hit@1}} \\
\midrule
UniIR (CLIP$\_$SF) & 9.6 & 9.8 & 10.5 & 8.2 & 11.5 & 11.0 & 0.1 & 0.1 & 0.1 & 0.0 & 0.1 & 23.7 & 9.2 & 8.5 & 20.5 & 21.3 & 9.0 \\
UniIR (BLIP$\_$SF) & 10.8 & 11.5 & 12.2 & 7.2 & 12.7 & 11.1 & 0.2 & 0.1 & 0.0 & 0.1 & 0.0 & 19.5 & 8.3 & 7.9   & 21.7 & 20.0  &  8.9 \\
VLM2Vec (2B) & 15.7 & 16.2 & 16.1 & 10.7 & 9.9 & 0.6 & 0.3 & 0.1 & 0.1 & 0.1 & 0.0 & 40.0 & 9.8 & 9.4 & 29.8 & 24.7 & 11.5\\
VLM2Vec (7B) & 35.1 & 38.0 & 42.5 & 5.4 & 17.3 & 10.0 & 0.4 & 0.4 & 0.1 & 0.1 & 0.1 & 33.5 & 16.8 & 16.4 & 44.6 & 48.2 & 19.3\\
VLM2Vec-v2 (2B) & 72.9 & 74.6 & 85.0 & 91.9 & 55.5 & 62.1 & 7.4 & 7.8 & 0.1 & 0.1 & 0.0 & 40.0 & 85.2 & 81.7 & 67.7 & \blue{100.0} & 53.3\\
LamRA (7B) & 17.2 & 16.7 & 17.4 & 22.1 & 26.0 & 15.2 & 9.9 & 9.9 & 0.2 & 6.8 & 3.3 & 42.2 & 44.6 & 51.7 & 45.4 & 62.0 & 24.4\\
GME (2B) & 67.1 & 66.2 & 67.9 & 80.3 & 27.7 & 52.7 & 2.3 & 3.7 & 0.1 & 0.0 & 0.1 & 47.6 & 67.5 & 63.7 & 89.9 & 99.7 & 46.0\\
GME (7B) & 79.0 & 77.8 & 89.3 & 87.2& 37.0 & 52.7 & 3.5 & 4.3 & 0.0 & 1.5 & 3.3 & 51.8 & 65.1 & 71.4 & 67.7 & \blue{100.0} &  49.5\\
\midrule
UniIR-FT (CLIP\_SF) & 55.8 & 56.1 & 58.9 & 47.3   & 50.2 & 56.8 & 6.4 & 4.7 & 3.3 & 3.6 & 2.5 & 42.9 & 37.3 & 45.6 & 61.8 & 53.3 & 36.6\\
UniIR-FT (BLIP\_SF) &  57.6 & 58.4 & 59.7  & 51.2 & 46.7 & 53.1 &  5.6 & 5.2 & 3.5 & 3.3 & 2.7 & 43.3  & 40.1 & 43.9 & 58.0 & 50.7 & 36.4 \\
\midrule
CodeMMR (2B) & \lblue{96.4} & \blue{97.4} & \blue{97.9} & \blue{99.5} & \blue{83.3} & \blue{84.9} & \blue{14.9} & \textbf{16.7} &  \blue{7.9} & \blue{7.4}  &  \blue{5.4} & \lblue{52.6} & \blue{93.5} & \lblue{94.3} & \blue{100.0} & \blue{100.0} & \blue{65.4} \\
CodeMMR (2B)-Mix & \blue{97.7} & \lblue{97.2} & \lblue{97.8} & \lblue{99.4} & \lblue{81.8} & \lblue{84.1} & \lblue{12.9} & \lblue{14.7} & \lblue{6.1} &  \lblue{6.2} & \lblue{5.8} &  \blue{52.8} & \lblue{93.2} & \lblue{94.5} & \blue{100.0} & \blue{100.0} &  \lblue{65.2} \\
\midrule
\multicolumn{18}{c}{\textit{Metric: nDCG@10}} \\
\midrule
UniIR (CLIP$\_$SF) & 18.6 & 19.1 & 20.2 & 15.3 & 25.3 & 22.9 & 0.2 & 0.3 & 0.2 & 0.1 & 0.2 & 44.2 & 14.7 & 15.3 & 30.1 & 27.6 & 15.9  \\
UniIR (BLIP$\_$SF) & 23.4 & 22.5 & 23.2 & 14.7 &  26.3 & 20.0 & 0.3 & 0.3 &  0.1 &0.1  & 0.1 & 38.2  & 13.8 & 14.2 & 27.9 & 26.9 &  15.8 \\
VLM2Vec (2B) & 31.0 & 30.7 & 30.4 & 16.0 & 22.1 & 2.3 & 0.6 & 0.4 & 0.2 & 0.2 & 0.2 & 49.3 & 14.1 & 12.2 & 37.9 & 34.9 &  17.7 \\
VLM2Vec (7B) & 50.3 & 50.0 & 56.4 & 11.2 & 30.0 & 19.1 & 2.0 & 2.5 & 0.3 & 0.4 & 0.2 & 53.1 & 32.3 & 35.7 & 52.0 & 55.0 & 28.2 \\
VLM2Vec-v2 (2B) & 83.7 & 82.2 & 90.8 & 95.2 & 69.3 & 75.5 & 8.8 & 9.3 & 0.3 & 0.2 & 0.3 & 40.9 & 88.4 & 87.4 & 99.7 & 95.2 & 58.0 \\
LamRA (7B) & 27.6 & 28.4 & 28.0 & 28.3 & 42.3 & 26.5 & 9.9 & 9.9 & 2.1 & 1.8 & 2.5 & 53.1 & 57.5 & 62.2 & \blue{100.0} & \blue{100.0} & 36.3 \\
GME (2B) & 73.1 & 70.8 & 78.0 & 87.9 & 44.9 & 68.7 & 3.2 & 4.3 & 0.3 & 0.6 & 0.8 & 58.1 & 72.4 & 68.8 & 93.0 & 99.8 & 51.5\\
GME (7B) & 82.3 & 81.6 & 93.0 & 92.1 & 54.5 & 69.3 & 5.0 & 5.9 & 3.0 & 8.5 & 3.7 & 43.0 & 71.1 & 73.6 & 74.1 & \blue{100.0} &  53.8 \\
\midrule
UniIR-FT (CLIP\_SF) & 72.2 & 68.7 & 67.8 & 57.4 & 59.7 & 67.0 & 8.8 & 7.9 & 5.4 & 5.2 & 3.0 & 46.9 & 40.7 & 51.2 & 63.9 & 61.2 & 42.9 \\
UniIR-FT (BLIP\_SF) & 69.9 & 70.2 & 68.2 & 63.5 & 55.6 & 64.1 & 7.4 & 8.0 & 4.9 & 5.1 & 3.2 & 48.1 & 44.2 & 50.0 & 60.7 & 59.3 &  42.7 \\
\midrule
CodeMMR (2B) & \blue{92.8} & \blue{90.5} & \blue{98.6} & \blue{99.8} & \blue{90.4} & \blue{91.5} & \blue{18.6} & \blue{19.7} & \blue{11.2} & \blue{12.3} & \blue{9.4} & \lblue{58.2} & \blue{97.8} & \blue{97.0} & \blue{100.0} & \blue{100.0} & \blue{68.0} \\
CodeMMR (2B)-Mix & \lblue{91.9} & \lblue{89.8} & \lblue{98.4} & \lblue{99.7} & \lblue{88.8} & \lblue{91.2} & \lblue{12.3} & \lblue{15.5} & \lblue{9.7} & 11.3 & 8.2 & \blue{59.5} & \lblue{93.2} & \lblue{94.5} & \blue{100.0} & \blue{100.0} & \lblue{66.5} \\
\bottomrule
\end{tabular}
}
\caption{\textbf{Performance on MMCoIR (Hit@1 and nDCG@10, \%):} Across model scales, CodeMMR substantially outperforms prior multimodal embedding models, highlighting the task’s difficulty and the need for specialized retriever learning. \blue{Blue} and \lblue{light blue} indicates the best and second best performance across the models.}
\label{tab:benchmark_results}
\end{table*}

\subsection{Dataset Structure}
\label{sec:dataset_structure}


MMCoIR employs a unified data schema to support a wide range of multimodal retrieval scenarios across vision–language–code modalities.
To ensure consistent query semantics across these heterogeneous datasets, we standardize how textual instructions are represented and interpreted during retrieval.

\noindent \textbf{Instruction formulation.} When datasets provide predefined textual instructions, we directly use them as queries. For datasets without explicit instructions, we introduce standardized prompts (e.g., \textit{Please retrieve the image that matches the description.''}) to clarify retrieval intent and explicitly define the input and output modalities. The interpretation of textual fields varies across datasets. In {ChartGen}, the text corresponds to chart descriptions, while in {ChartEdit} and {DiagramGenBenchmark}, it represents editing or repair instructions (e.g., \textit{Change the line color to red.''}). Notably, {ChartEdit} and {DiagramGenBenchmark} also introduce several novel retrieval settings, such as text + code $\rightarrow$ image, text + code $\rightarrow$ image + code, and text + image $\rightarrow$ code, which are absent from the training data and therefore serve as unseen tasks for evaluating model generalization.

\noindent \textbf{Sampling and balance.} To ensure balanced representation across domains, we subsample instances from larger datasets to maintain comparable scales. In most cases, the training set contains 100k instances and the test set contains 2k instances. Some datasets include multiple subsets for finer granularity. For example, {MMSVG} is divided into {MMSVG-Icon} and {MMSVG-Illustration}, each containing 100k training and 2k test instances, while {PlantUML} is partitioned into {PlantUML-Act} and {PlantUML-Seq}, focusing on activity and sequence diagrams, respectively. Datasets, such as {Sketch2Code}, {ChartEdit}, and {DiagramGenBenchmark}, only provide test data and are therefore used to assess the model’s generalization ability on unseen domains.

\section{Experiments}

Leveraging MMCoIR, we evaluate existing vision–language embedding models on the benchmark and train a unified code–image–text embedding model.

\paragraph{Baselines.} We evaluate our method against a diverse set of state-of-the-art vision--language representation and retrieval models to ensure a fair and comprehensive comparison across domains. The baselines include two UniIR~\cite{wei2024uniir} variants, UniIR (CLIP\_SF) and UniIR (BLIP\_SF), which are unified multimodal retrievers trained with instruction tuning to support diverse retrieval formats.
We also compare with several VLM-derived embedding models: GME~\cite{zhang2024gme} (2B and 7B), VLM2Vec~\cite{jiang2025vlmvec} (2B and 7B), VLMVec-v2~\cite{meng2025vlm2vec}, and LamRA~\cite{liu2025lamra}---which are primarily fine-tuned from the Qwen2VL~\cite{wang2024qwen2} backbone on large-scale image--text data. These models extend Qwen2VL’s multimodal capabilities by incorporating task-specific objectives and instruction-aware embeddings. Finally, we fine-tune the UniIR models on the MMCoIR training set (denoted as UniIR-FT) to better adapt them to our tasks.
For more details about the baselines, please refer to the supplementary material.

\paragraph{Implementation Details.} CodeMMR is implemented using the PyTorch framework and built upon the Qwen2-VL-2B-Instruct backbone~\cite{wang2024qwen2}. We employ a contrastive learning objective for multimodal alignment following the VLM2Vec setup~\cite{jiang2025vlmvec}. The model is fine-tuned with LoRA ($r{=}8$) in bf16 precision, using EOS pooling with $\ell_2$ normalization and a contrastive loss temperature of 0.02. All experiments are conducted on 8 NVIDIA A100 80GB GPUs with a per-device batch size of 64 (global 512).
Input text sequences are truncated to a maximum length of 256 tokens, which is consistent with most existing multimodal embedding models, ensuring fair comparison and stable convergence. Optimization is performed with AdamW (learning rate $5\times10^{-5}$, 100 warmup steps, 1000 total steps) under a linear learning-rate scheduler. The vision encoder is frozen, and only language model parameters are updated via LoRA. The entire training completes in about 30 hours, and unless otherwise specified.

\begin{table}[t]
\centering
\setlength{\tabcolsep}{3pt}
\resizebox{\linewidth}{!}{
\begin{tabular}{l|c|cc|cccc}
\toprule
\multirow{3}{*}{\textbf{\large Methods}} & 
\textbf{\large WebUI} & 
\multicolumn{2}{c|}{\textbf{\large Data Charts}} & 
\multicolumn{4}{c}{\textbf{\large Schematic Diagram}} \\
\cmidrule(lr){2-2} \cmidrule(lr){3-4} \cmidrule(lr){5-8}
& \textbf{Sketch2Code} & 
\multicolumn{2}{c|}{\textbf{ChartEdit}} & 
\multicolumn{4}{c}{\textbf{DiagramGenBenchmark}} \\
\cmidrule(lr){2-2} \cmidrule(lr){3-4} \cmidrule(lr){5-8}
& $q_i \to r_c$ & 
$q_c \to r_i$ & $q_{t,c} \to r_i$ & 
$q_c \to r_i$ & $q_i \to r_c$ & $q_t \to r_c$ & $q_{t,i} \to r_c$ \\
\midrule
\multicolumn{8}{c}{\textit{Metric: hit@1}} \\
\midrule
UniIR (BLIP\_SF) & 0.0 & 0.5 & 0.4 & 15.5 & 17.9 & 0.1 & 20.1 \\
UniIR (CLIP\_SF) & 0.0 & 0.6 & 0.4 & 16.3 & 18.2 & 0.2 & 21.2 \\
VLM2Vec (2B) & \lblue{0.7} & 1.9 & 2.8 & 21.5 & 24.8 & 0.3& 25.5 \\
VLM2Vec (7B) & 0.5 & 10.7 & 8.9 & 32.3 & 36.2 & 0.5 & 32.1 \\
VLM2Vec-v2 (2B) & 0.1 & 88.7 & 15.7 & 85.9 & 87.4 & 0.4 & 85.0 \\
LamRA (7B) & 0.7 & 28.4 & 11.0 & 70.0 & 52.2 & 0.4 & 52.0 \\
GME (2B) & 0.4 & 85.9 & 13.2 & 72.6 & 72.2 & 0.7 & 84.0 \\
GME (7B) & \blue{2.7} & 89.0 & 15.4 & 46.3 & 82.2 & 0.4 & 87.5 \\

\midrule
UniIR-FT (BLIP\_SF) & {0.6} & 27.9 & 10.3 & 69.1 & 51.0 & 2.3 & 51.2 \\
UniIR-FT (CLIP\_SF) & 0.5 & 28.0 & 10.5 & 70.2 & 51.7 & 1.9 & 52.0 \\
\midrule

CodeMMR & 0.5 & \blue{100.0} & \lblue{27.1} & 94.8 & \blue{94.4} & \lblue{8.6} & \blue{88.0} \\
CodeMMR-Mix (2B) & \lblue{0.7} & \blue{100.0} & \blue{27.3} & \blue{95.2} & \lblue{93.7} & \blue{8.7} & \blue{89.0} \\
\midrule
\multicolumn{8}{c}{\textit{Metric: nDCG@10}} \\
\midrule

UniIR (BLIP\_SF) & 0.0 & 1.6 & 1.3 &  25.2 & 26.7 & 1.2 &  28.4\\
UniIR (CLIP\_SF) & 0.1 & 1.8 & 1.3 & 26.6 & 25.3 & 1.3 & 30.3 \\
VLM2Vec (2B) & \blue{2.1} & 6.6 & 6.3 & 29.9 & 30.6 & 1.7 & 36.4 \\
VLM2Vec (7B) & 1.2 & 27.5 & 31.2 & 55.2 & 60.1 & 1.7 & 63.8 \\
VLM2Vec-v2 (2B) & 1.1 & 95.4 & 32.5 & 91.7 & 92.4 & 1.6 & 91.5 \\
LamRA (7B) & 1.5 & 40.0 & 26.1 & 81.5 & 65.2 & 1.7 & 63.8 \\
GME (2B) & 1.3 & 90.8 & 28.7 &81.7 & 79.4 & 1.8 & 91.4 \\
GME (7B) & 1.3 & 93.8 & 31.2 & 61.5 & 84.9 & 1.7 & 93.5 \\

\midrule
UniIR-FT (BLIP\_SF) & 1.4 & 39.2 & 26.4 & 82.3 & 66.1 & 5.6 & 64.5 \\
UniIR-FT (CLIP\_SF) & 1.2 & 38.8 & 26.8 & 82.9 & 67.0 & 4.7 & 65.6 \\
\midrule

CodeMMR (2B) & 1.5 & \blue{100.0} & \blue{43.2} & \lblue{97.8} & \blue{97.3} & \blue{17.1} & \lblue{93.7} \\
CodeMMR-Mix (2B) & \lblue{1.7} & \blue{100.0} & \lblue{43.1} & \blue{97.9} & \lblue{97.1} & \lblue{16.8} & \blue{95.7} \\
\bottomrule
\end{tabular}
}
\caption{Performance comparison on unseen datasets.}
\label{tab:benchmark_results_additional}
\end{table}

\section{Performance over Retrieval Tasks}
Table~\ref{tab:benchmark_results} shows results on seen datasets, whereas Table~\ref{tab:benchmark_results_additional} covers \emph{unseen} domains and novel tasks, allowing comparison between \emph{in-} and \emph{out-of-distribution} performance. We summarize the main findings as follows:

\noindent \textbf{i) Fine-tuning on MMCoIR substantially boosts retrieval performance across all datasets.}
In Table~\ref{tab:benchmark_results}, CodeMMR (2B) achieves the best overall performance on datasets with training data, demonstrating its strong multimodal representation learning when exposed to supervised alignment signals. Fine-tuning on MMCoIR notably improves all baseline models, confirming the dataset’s effectiveness in enhancing cross-modal understanding. Even for relatively weaker baselines such as UniIR, fine-tuning leads to large gains, over 20 points in \textit{Hit@1} on average, highlighting the benchmark’s value as a unifying multimodal corpus. Among all approaches, CodeMMR surpasses the best fine-tuned baselines by a wide margin, reaching 65.4 \textit{Hit@1}, which demonstrates its ability to integrate code reasoning with visual and textual semantics.
\begin{figure}[!t]
    \centering
    \includegraphics[width=0.8\linewidth]{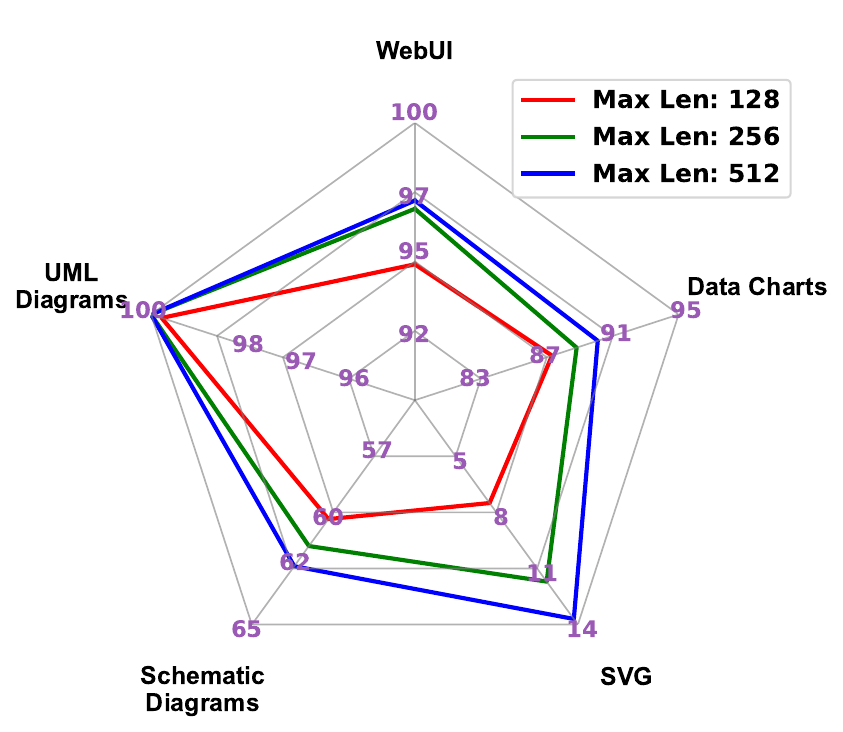}
    \caption{Impact of maximum input length during training on retrieval performance.}
    \label{fig:len}
\end{figure}

\begin{figure*}
    \centering
    \includegraphics[width=\linewidth]{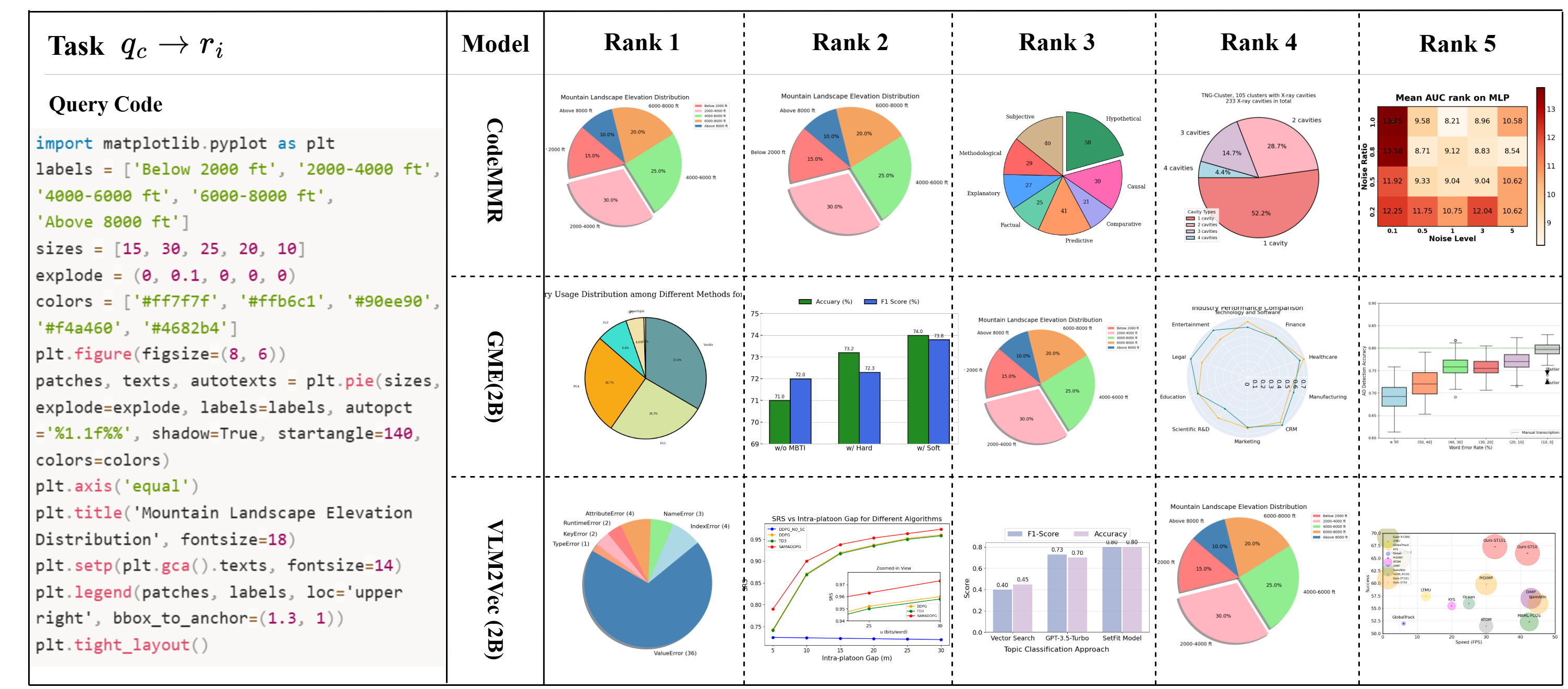}
    \caption{RAG by CodeMMR consistently boosts MLLMs on code generation, including execution success and structural fidelity on {ChartMimic} and {WebCode2M}.}
    \label{fig:examples}
\end{figure*}

\noindent \textbf{ii) Performance varies significantly across domains due to inherent differences in visual and structural complexity.}
As illustrated in Table~\ref{tab:benchmark_results}, UML diagrams yield the highest retrieval accuracy (\textit{Hit@1 = 100\%}), likely because their code snippets are short, structurally simple, and rely heavily on symbolic text matching. In contrast, SVG graphics represent the most challenging domain: SVG code is long, compositional, and geometrically complex, making the mapping between visual content and code tokens highly non-trivial (see Appendix). While image-to-code retrieval in SVG remains difficult, code-to-image directions perform relatively better, suggesting asymmetric modality alignment. WebUI also achieves notably strong results ($>97\%$ \textit{Hit@1}), possibly benefiting from abundant high-quality HTML/CSS data used in pretraining and the strong textual grounding of web layout descriptions.

\noindent \textbf{iii) CodeMMR shows noticeable performance improvements on unseen datasets and novel retrieval tasks} As shown in Table~\ref{tab:benchmark_results_additional}, the model performs consistently better than baseline approaches across various unseen settings. On the DiagramGen Benchmark, CodeMMR demonstrates good cross-domain transfer, maintaining competitive accuracy despite the lack of task-specific supervision. In ChartEdit, it retrieves chart images from plotting code effectively, though performance decreases when handling textual edit instructions, indicating that compositional reasoning over multimodal inputs remains challenging. On Sketch2Code, where inputs are human-drawn sketches of web interfaces, only limited improvement is observed, reflecting the inherent difficulty in bridging the visual gap between sketches and rendered layouts. Overall, these results indicate that CodeMMR can generalize reasonably well to new domains, while there is still room for improvement on abstract and instruction-driven retrieval scenarios.


\subsection{Ablation Studies}
\noindent \textbf{iv) Conventional multimodal retrieval datasets offer little benefit on MMCoIR.}
A common practice in retriever training is to combine general-purpose and domain-specific corpora to improve generalization~\cite{liu2025codexembed}, which often outperforms training solely on specialized data. To test this on MMCoIR, we trained \textit{CodeMMR-Mix} by incorporating the large-scale training datasets from {VLM2Vec}~\cite{jiang2025vlmvec}.
As shown in Table~\ref{tab:benchmark_results}, CodeMMR-Mix achieves results comparable to, or slightly worse than, CodeMMR across most domains, with marginal differences in \textit{Hit@1} and \textit{nDCG@10}.
Moreover, on unseen domains such as ChartEdit and DiagramGenBenchmark, CodeMMR-Mix shows no measurable improvement, indicating that the inclusion of traditional multimodal retrieval data does not enhance the model’s generalization ability.
These findings suggest that MMCoIR alone provides sufficiently rich cross-modal and cross-domain supervision for unified code retrieval.

\noindent \textbf{v) Longer training input length leads to consistent performance gains across domains.} 
Most existing multimodal embedding models are trained with a default maximum input length of 256 tokens~\cite{jiang2025vlmvec,meng2025vlm2vec}, which is often inadequate for structured programming language like SVG XML,  causing truncation of critical semantic or syntactic elements. 
To evaluate the influence of input length during training, we fine-tuned {CodeMMR} with maximum sequence lengths of 128, 256, and 512 tokens. As shown in Figure~\ref{fig:len}, extending the input length consistently improves retrieval accuracy across all domains, especially in structure-heavy tasks like SVG (from 7.5 to 13.7) and schematic diagrams (from 60.3 to 62.4). These findings demonstrate that allowing longer training input sequences enhances the model’s capacity to encode fine-grained structural and cross-modal dependencies
.

\subsection{Qualitative Analysis}  Figure~\ref{fig:examples} presents a qualitative comparison of retrieval results for the code-to-image task, where the input query is a Python matplotlib script describing a pie chart of mountain elevation distribution. As illustrated, our CodeMMR model retrieves the most relevant visualization at Rank 1, precisely matching the intended structure, labels, and color composition specified in the code. In contrast, GME (2B) and VLMVec (2B) often retrieve figures with different semantic contexts or chart types (e.g., bar charts, radar plots), failing to capture the fine-grained visual correspondence implied by the plotting commands. These results highlight CodeMMR’s superior capability in aligning executable code semantics with the corresponding visual representation, demonstrating its robust understanding of multimodal compositional structures.

\begin{figure}
    \centering
    \includegraphics[width=\linewidth]{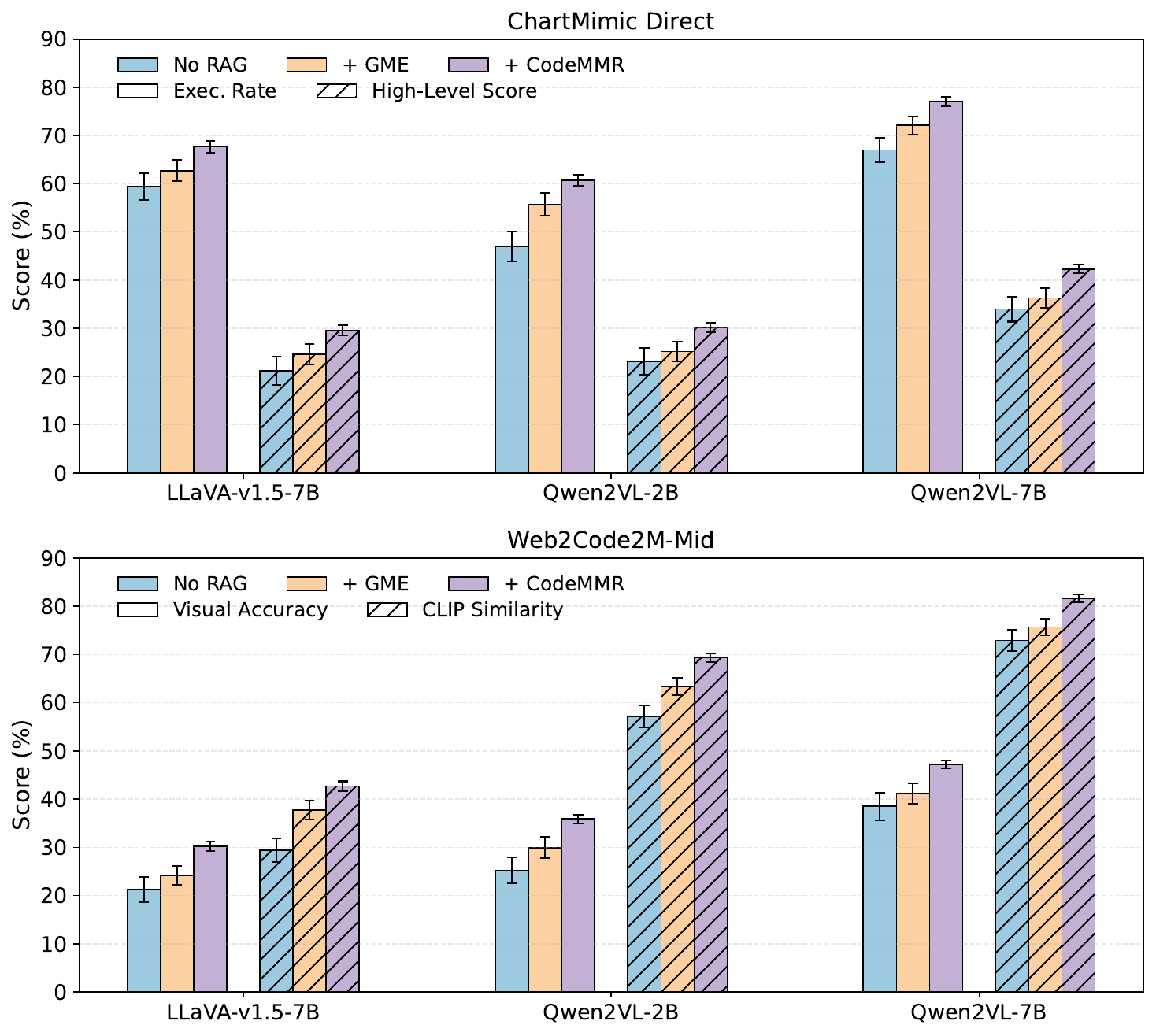}
    \caption{Comparison of retrieval-augmented code generation across three VLMs on ChartMimic Direct and Web2Code2M-Mid.}
    \label{fig:rag_comparison}
\end{figure}

\section{Multimodal Code-RAG with CodeMMR}
To further investigate whether CodeMMR can enhance code reasoning and generation, we evaluate its RAG performance on two representative benchmarks of image-to-code generation: \textit{ChartMimic Direct}~\cite{yang2025chartmimic} and \textit{WebCode2M-Mid}~\cite{gui2025webcode2m}.
To prevent data contamination, retrieval is performed exclusively within the code corpora of the \textit{ChartGen} and \textit{WebSight} training subsets from MMCoIR, ensuring that no validation samples overlap. We follow the original experimental configurations of these benchmarks to ensure comparability with prior results. We adopt \textit{LLaVA-7B}, \textit{LLaVA-13B}, and \textit{Qwen2VL-7B} as backbone MLLMs, and compare their performance under three configurations: \textit{No RAG}, \textit{+ GME (7B)}, and \textit{+ CodeMMR}.

Across both benchmarks, {CodeMMR} consistently improves generation quality, outperforming both non-retrieval and GME-based retrieval settings. On \textit{ChartMimic Direct}, it yields notable gains over the No-RAG baseline, with an average improvement of {+10.0$\uparrow$ Execution Rate} and {+7.6$\uparrow$ High-Level Score}, surpassing GME by an additional 4--5 points. Similarly, on {WebCode2M-Mid}, where evaluation focuses on Visual Accuracy and CLIP Similarity, CodeMMR achieves the largest boost, improving scores by {+9.4$\uparrow$} and {+10.8$\uparrow$}, respectively. For instance, {Qwen2VL-2B} improves from 25.2 to 35.9 in Visual Accuracy and from 57.2 to 69.4 in CLIP Similarity, while {Qwen2VL-7B} increases from 38.5 to 47.2 and from 72.9 to 81.7. These consistent improvements indicate that the retrieved code snippets from CodeMMR provide more reliable structural and stylistic priors, helping MLLMs generate code that is both visually accurate and executable.


\begin{figure}
    \centering
    \includegraphics[width=\linewidth]{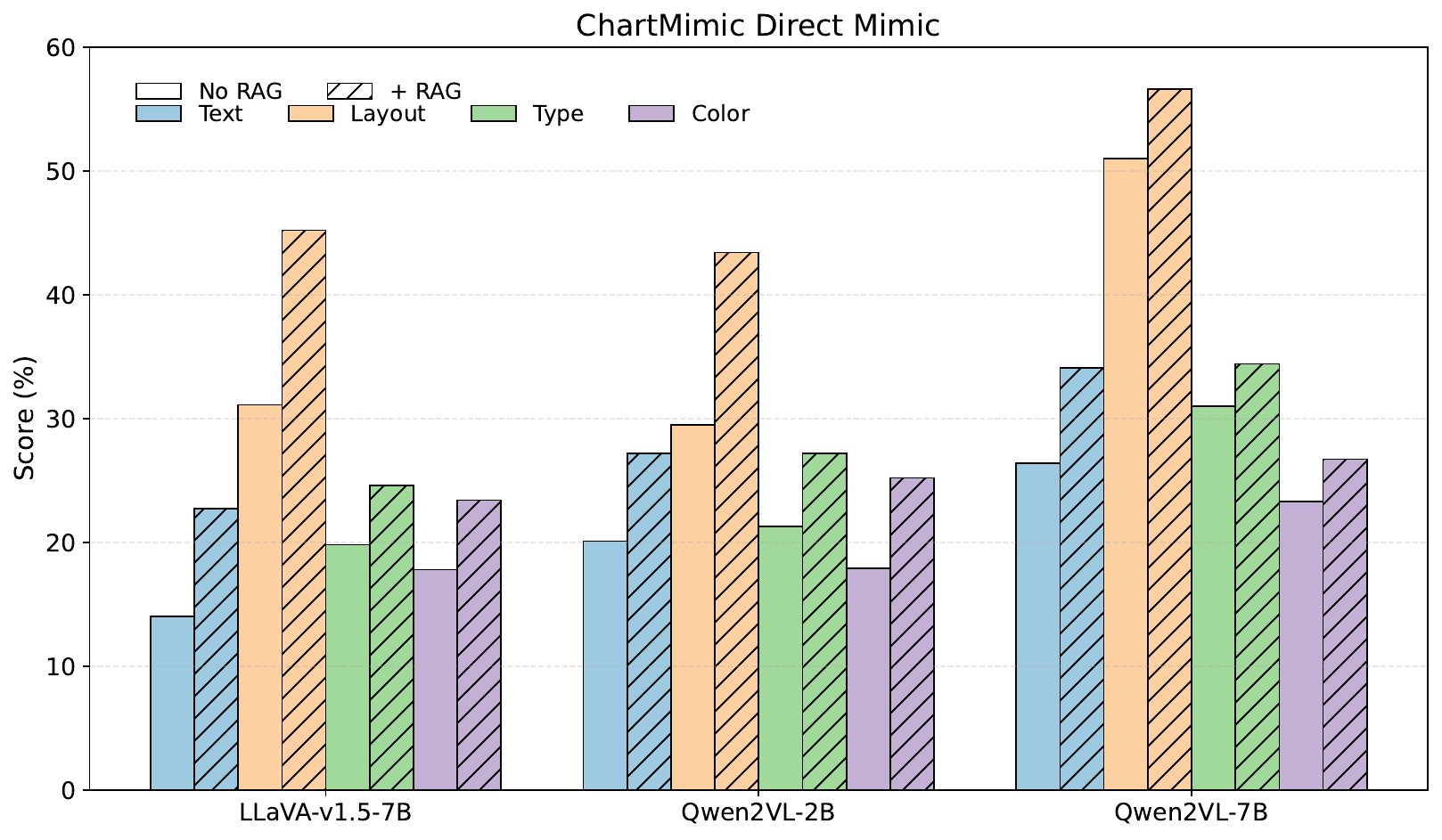}
    \caption{Low-level visual metric comparison on ChartMimic Direct before and after applying CodeMMR retrieval augmentation. Higher values indicate better performance.}
    \label{fig:chartmimic_scores}
\end{figure}


\paragraph{Qualitative analysis} 

Beyond aggregate metrics, we conduct a fine-grained evaluation on \textit{ChartMimic Direct} to analyze how retrieval augmentation improves different aspects of visual code generation. Following~\cite{yang2025chartmimic}, we evaluate four low-level metrics: \textit{Text}, \textit{Layout}, \textit{Type}, and \textit{Color}, which measure the accuracy of textual elements, spatial layout, chart type, and color fidelity. As shown in Fig.~\ref{fig:chartmimic_scores}, CodeMMR consistently improves performance across all dimensions. For example, in Qwen2VL-7B, the scores increase from 26.4\%, 51.0\%, 31.0\%, and 23.3\% to 34.1\%, 56.6\%, 34.4\%, and 26.7\%, respectively. Similar trends are observed for LLaVA-v1.5-7B and Qwen2VL-2B, with text and layout scores improving by about 7--14\%. These results confirm that retrieval augmentation enhances both semantic accuracy and structural fidelity in code generation.


To better understand these improvements, we manually inspect 200 chart–code pairs from the Qwen2VL-7B runs. Retrieval augmentation primarily mitigates two dominant error types. (i) \textit{textual and labeling mistakes}, such as missing axis titles or incorrect legend names, occur less frequently because retrieved exemplars provide aligned chart descriptions that guide precise text rendering. Second, \textit{layout inconsistencies}, including misplaced legends or distorted subplot arrangements, are reduced since the retrieved code offers robust structural templates (e.g., subplot definitions and legend positioning). Overall, CodeMMR retrieval grounds generation on realistic chart implementations, reducing hallucinations and improving both semantic coherence and structural alignment in code synthesis.

\section{Conclusion}
Visual information is a significant component of intelligence, for both humans and machines.
In this work, we construct a multimodal retrieval benchmark MMCoIR to better evaluate the real-world and diverse retrieval, including code, images, and texts.
By benchmarking existing models, we find the task is challenging.
Therefore, we presented CodeMMR, a unified multimodal and retrieval model that bridges natural language, code, and images within a shared embedding space.
Extensive experiments demonstrate that CodeMMR achieves state-of-the-art performance and substantially enhances retrieval-augmented code generation. 
Together with the MMCoIR benchmark, our work establishes the first comprehensive framework for evaluating and advancing multimodal code retrieval across diverse visual and linguistic domains.
We plan large-scale pretraining, reasoning-intensive, and the extension to more modalities, e.g., video, as future directions.

\section*{Acknowledgements} This work has received funding from the European Union’s Horizon Europe research and innovation programme project TrustLLM under grant agreement No 101135671.
\newpage
{
    \small
    \bibliographystyle{ieeenat_fullname}
    \bibliography{main}
}

\appendix


\clearpage
\setcounter{page}{1}
\maketitlesupplementary


\section{Dataset Statistics}

\noindent \textbf{Data Schema.}
Each training instance consists of six keys: \texttt{qry}, \texttt{qry\_img\_path}, \texttt{pos\_text}, \texttt{pos\_img\_path}, \texttt{neg\_text}, and \texttt{neg\_img\_path}. The \texttt{qry} field contains the structured textual component (natural language and/or code), while \texttt{qry\_img\_path} specifies the associated image if available. Positive and negative examples are defined through their respective text and image paths, enabling contrastive multimodal learning. Evaluation instances follow a similar structure with four keys: \texttt{qry\_text}, \texttt{qry\_img\_path}, \texttt{tgt\_text}, and \texttt{tgt\_img\_path}. Different retrieval directions are instantiated by populating the appropriate modalities, such as image-to-code ($q_i \rightarrow r_c$), code-to-image ($q_c \rightarrow r_i$), or text-to-code ($q_t \rightarrow r_c$). The special \texttt{[image]} token in text fields explicitly marks multimodal inputs, ensuring consistent processing across all retrieval configurations. Comprehensive dataset statistics are reported in Table~\ref{tab:datasets}.

\begin{figure*}[!t]
    \centering
    \includegraphics[width=\textwidth]{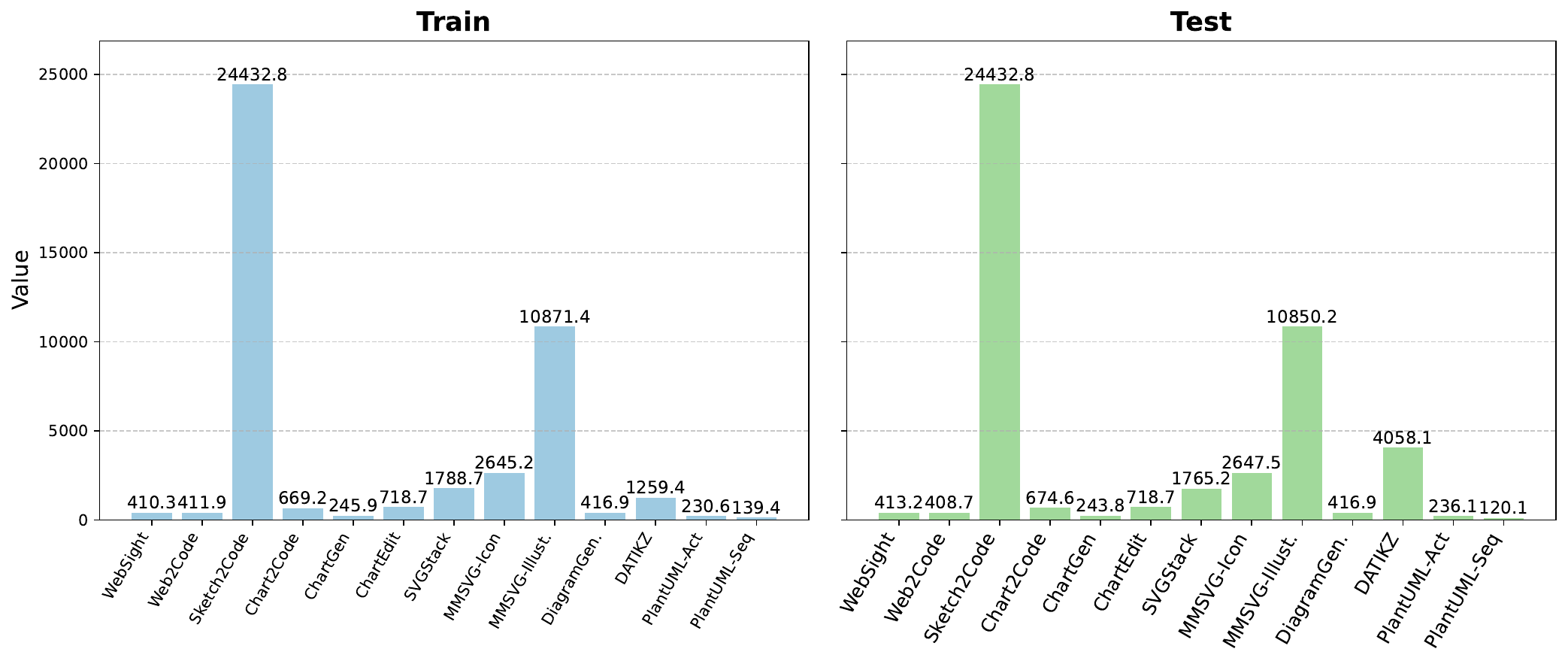}
    \caption{Average token length distribution across datasets in train/test splits.}
    \label{fig:length}
\end{figure*}

\noindent \textbf{Dataset Token Length Analysis.} Figure~\ref{fig:length} presents the average token lengths of training and test samples across all datasets. We observe that the token-length distributions of the train and test splits are highly consistent, indicating no significant distributional drift between them. Most datasets have moderate average lengths around several hundred tokens, suggesting that they can be efficiently processed under typical model context limits. However, a few datasets exhibit exceptionally long sequences: Sketch2Code, MMSVG-Illustration, and MMSVG-Icon contain substantially longer textual representations (over 10k tokens on average in Sketch2Code and MMSVG-Illustration). Such long sequences pose challenges for multimodal retrieval models when the token budget is limited (e.g., 256 or 512 tokens), potentially leading to information truncation and degraded retrieval performance. Interestingly, despite some datasets such as WebSight, Web2Code, and Chart2Code also having average lengths slightly beyond the 256-token range, our retrieval model maintains strong performance on these sources. This indicates that our training strategy enables robust generalization even when facing inputs that exceed the nominal token limit. The model effectively captures the cross-modal alignment without overfitting to input length, suggesting that the learned representation scales gracefully with content complexity.

\begin{figure}[!t]
    \centering
    \includegraphics[width=\linewidth]{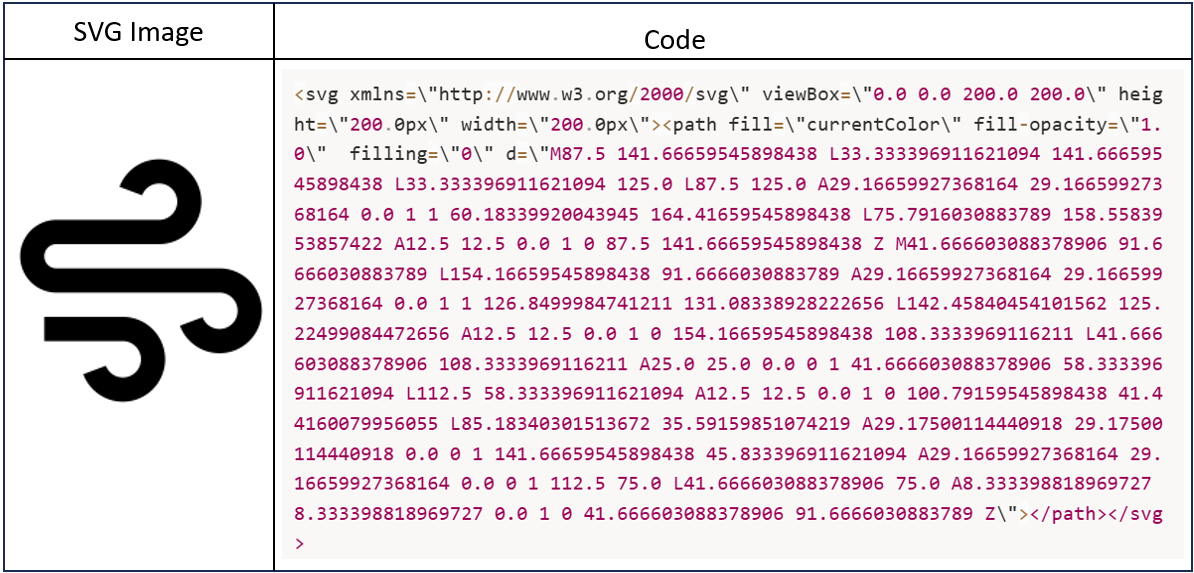}
    \caption{SVG example.}
    \label{fig:svg}
\end{figure}

\noindent \textbf{SVG Example.} As shown in Figure~\ref{fig:svg}, we illustrate an example of an SVG image and its corresponding XML‐based code representation. The SVG format encodes visual shapes using textual geometric primitives such as paths, curves, and color attributes. Although the image and its underlying code describe the same content, their modalities differ drastically — the visual domain emphasizes spatial patterns and holistic structure, whereas the XML text focuses on low-level coordinate and syntax tokens. This discrepancy makes it extremely challenging for multimodal retrieval systems to establish direct correspondence between visual and textual cues. The mapping from complex XML sequences to perceptually coherent image features is highly non-linear, and minor code variations may yield visually indistinguishable results. Consequently, learning effective cross-modal alignment between SVG code and rendered images remains a difficult problem, limiting retrieval accuracy and representation consistency in multimodal tasks.

\section{Additional Results}
\begin{figure}[H]
    \centering
    \includegraphics[width=0.9\columnwidth]{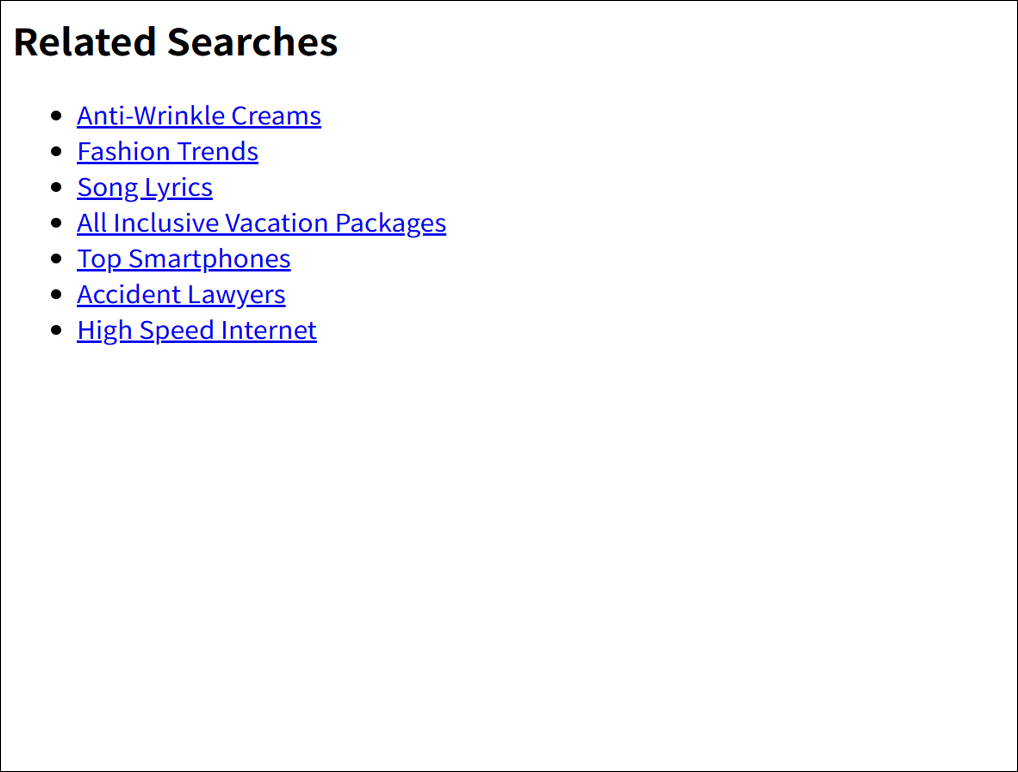}
    \caption{Qualitative comparison of web page code generation on WebCode2M-Mid: LLaVA-7B.}
    \label{fig:results_7b}
\end{figure}

\begin{figure}[H]
    \centering
    \includegraphics[width=0.9\columnwidth]{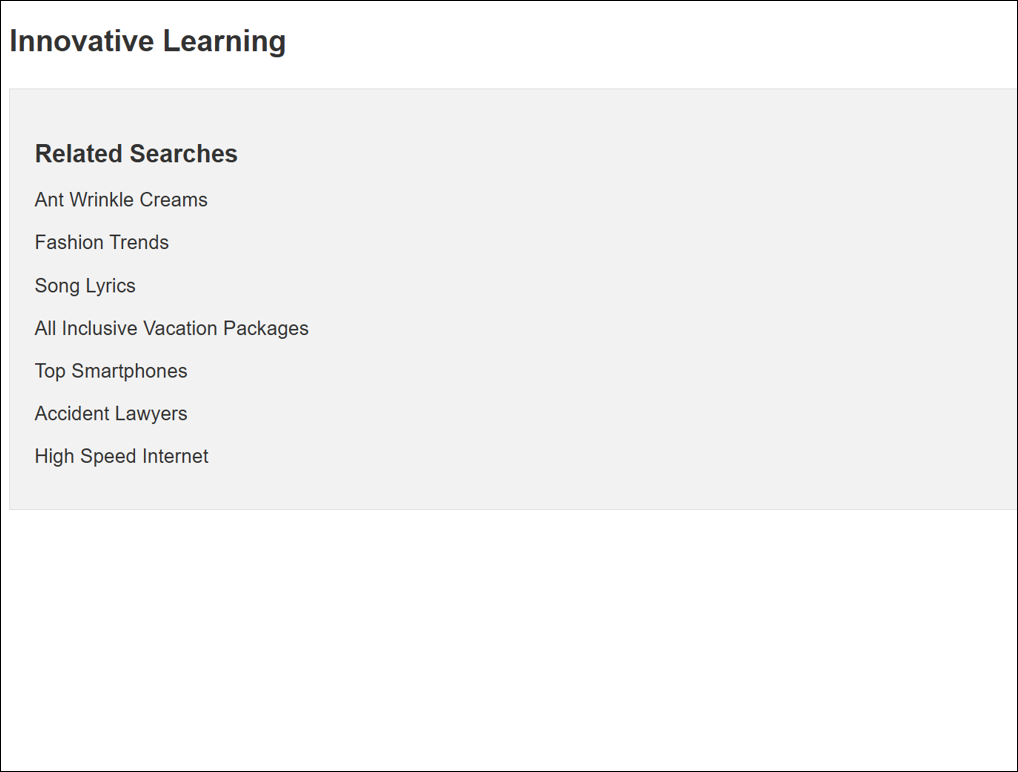}
    \caption{Qualitative comparison of web page code generation on WebCode2M-Mid: LLaVA-13B.}
    \label{fig:results_13b}
\end{figure}

\begin{figure}[H]
    \centering
    \includegraphics[width=0.9\columnwidth]{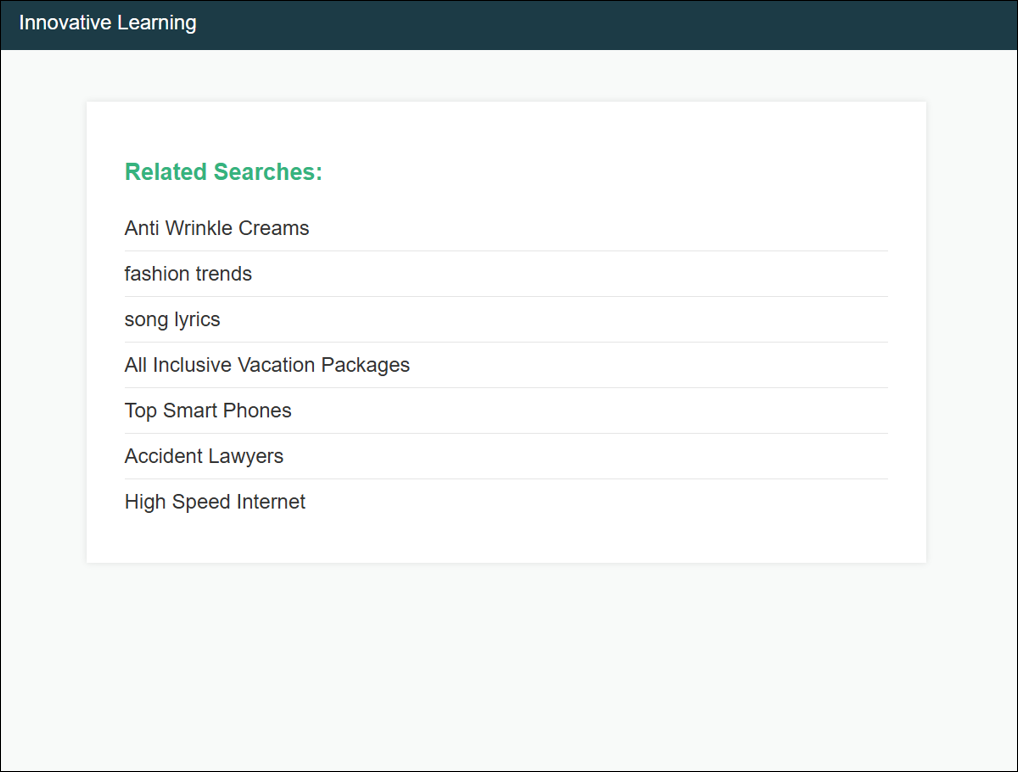}
    \caption{Qualitative comparison of web page code generation on WebCode2M-Mid: LLaVA-13B + CodeMMR.}
    \label{fig:results_13b_rag}
\end{figure}

\begin{figure}[H]
    \centering
    \includegraphics[width=0.9\columnwidth]{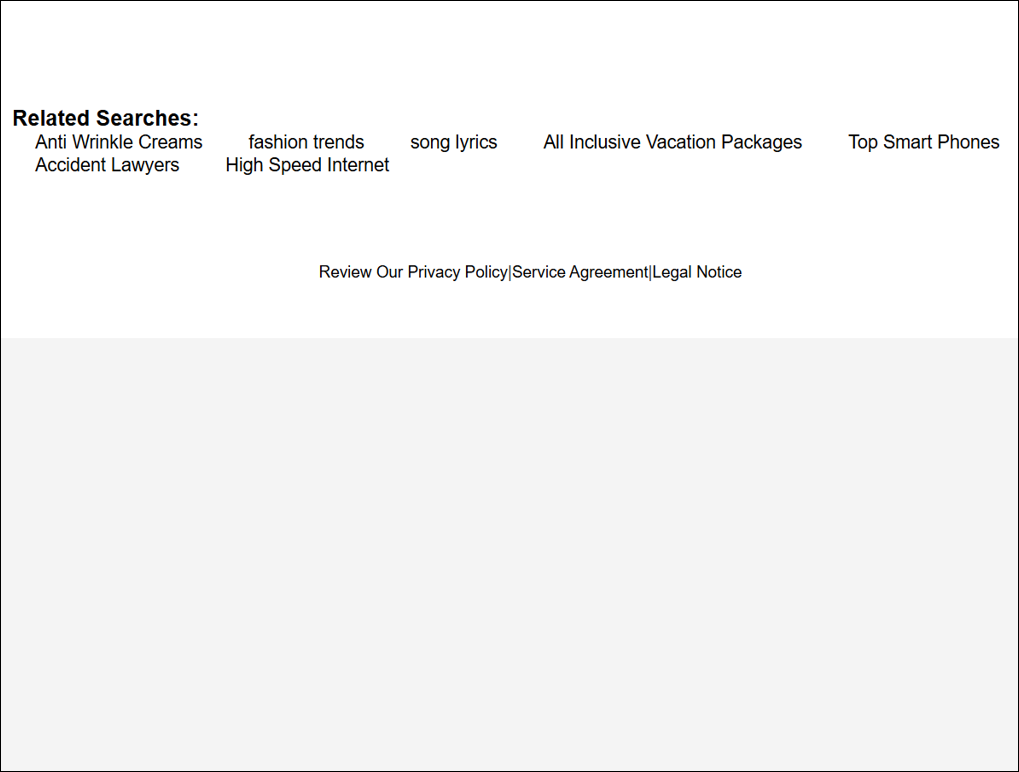}
    \caption{Qualitative comparison of web page code generation on WebCode2M-Mid: WebCoder.}
    \label{fig:results_webcoder}
\end{figure}

\begin{figure}[H]
    \centering
    \includegraphics[width=0.9\columnwidth]{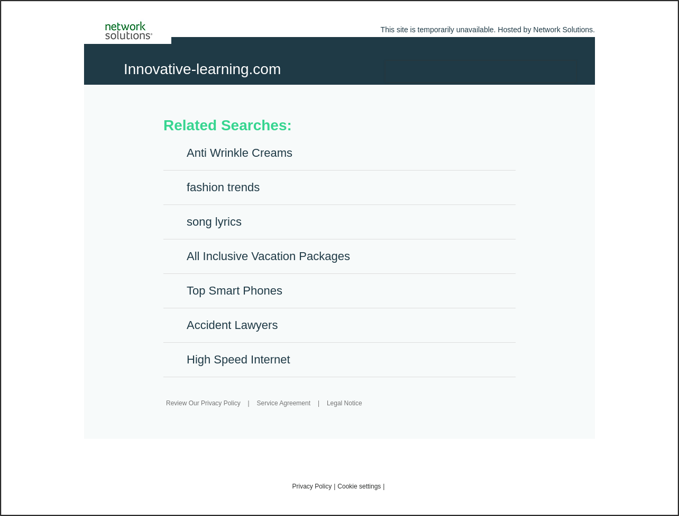}
    \caption{Qualitative comparison of web page code generation on WebCode2M-Mid: Ground Truth.}
    \label{fig:results_gt}
\end{figure}

\paragraph{Multi-modal code RAG exhibits stronger robustness and better performance over fine-tuned VLM.}
We evaluate multiple vision-language models on the task of generating webpage code from a given screenshot on the WebCode2M-Mid dataset.
Figure~\ref{fig:results_7b} shows the result of \textbf{LLaVA-7B}, which captures the basic list structure and hyperlink styling, but completely misses the header bar and overall page layout, rendering only a plain bulleted list without any container or background styling.
Figure~\ref{fig:results_13b} presents \textbf{LLaVA-13B}, which correctly reproduces the header and the search item list within a styled container, but loses hyperlink formatting and applies an overly plain gray background, with no color differentiation for the section title.
Figure~\ref{fig:results_13b_rag} illustrates \textbf{LLaVA-13B + CodeMMR (ours)}, which integrates RAG with a code-aware multimodal retriever. The generated page closely replicates the dark header bar, the teal-colored ``Related Searches'' title, and the vertically spaced list items, achieving the highest visual fidelity among all evaluated models.
Figure~\ref{fig:results_webcoder} shows \textbf{WebCoder}~\cite{gui2025webcode2m}, a ViT-based model fine-tuned on WebCode2M, which recovers the section title and list items but renders them in a horizontally misaligned layout, losing the vertical list structure and introducing irrelevant footer text.
Figure~\ref{fig:results_gt} presents the \textbf{Ground Truth}, i.e., the original rendered webpage, featuring a dark navigation header, a teal ``Related Searches'' heading, and a clean vertically stacked list of search terms with proper spacing and footer links.
Overall, our method LLaVA-13B + CodeMMR achieves the best reconstruction quality, closely matching the ground truth in layout, typography, and hierarchical organization, demonstrating the effectiveness of retrieval-enhanced multimodal reasoning for code generation from web screenshots.


\end{document}